\documentclass[a4paper]{article}
\usepackage{graphicx}
\begin{document}
\title{Recent Highlights from the PHENIX Heavy-Ion Program at RHIC}

\author{{\slshape Baldo Sahlmueller$^1$, for the PHENIX collaboration}\\[1ex]
$^1$ Goethe University, Max-von-Laue-Str. 1, 60438 Frankfurt, Germany}


\date{ }
\maketitle

\begin{abstract}
Over the last decade it has been established that a quark-gluon plasma (QGP) is formed in ultrarelativistic A+A collisions at RHIC energies. In recent years, detector upgrades have enabled the detailed study of this hot and dense matter. Important probes, among others, are direct photons and heavy flavor observables. Although the RHIC d+Au program was originally undertaken to study initial state and cold nuclear matter effects, recent measurements at both RHIC (d+Au) and the LHC (p+Pb) have found evidence for collective phenomena in these small systems.
\end{abstract}

\section{Direct Photons}

\begin{figure}[ht]
\begin{minipage}[t]{55mm}
\centerline{\includegraphics[width=\textwidth]{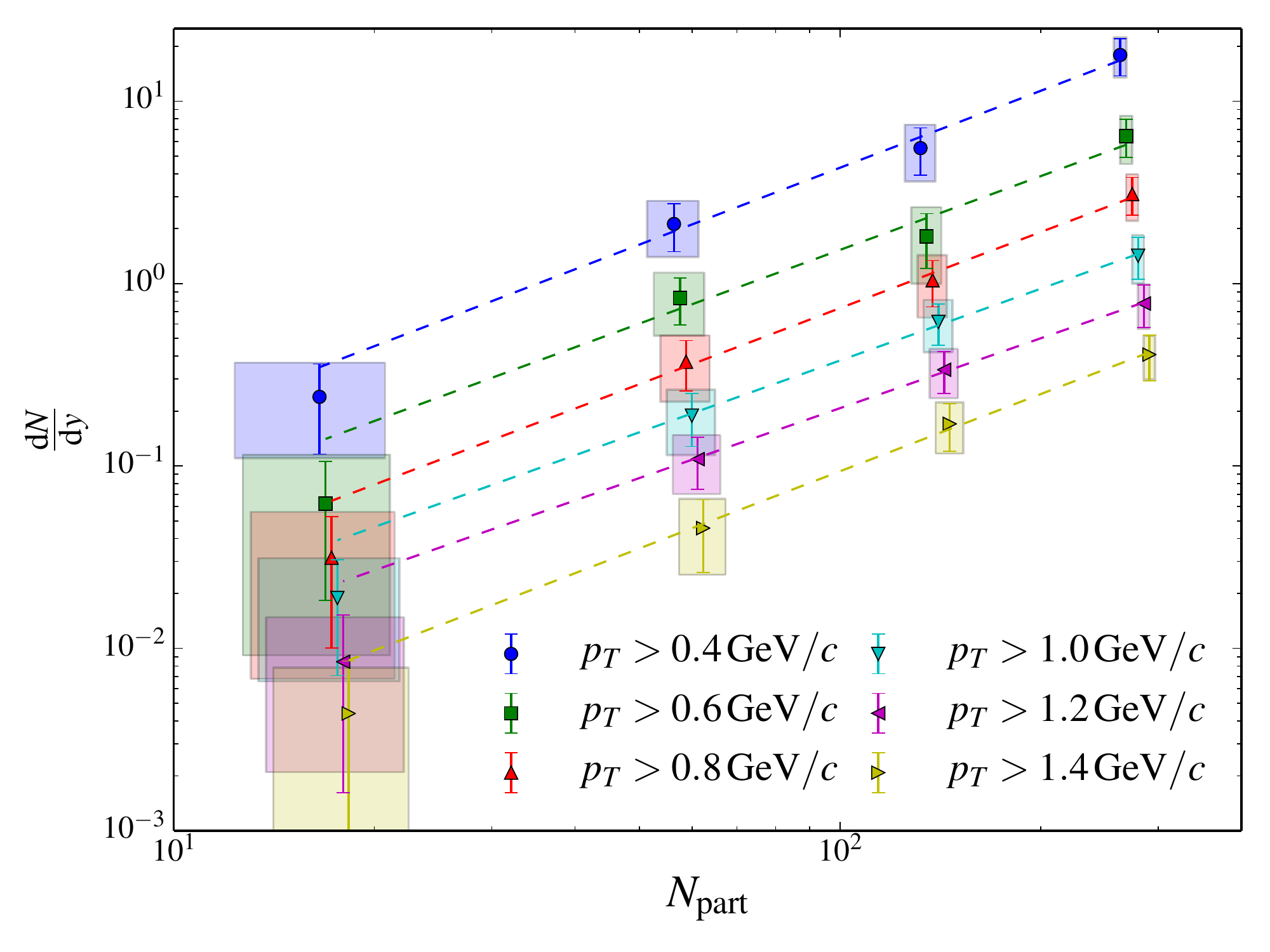}}
\end{minipage}
\hspace{\fill}
\begin{minipage}[t]{85mm}
\centerline{\includegraphics[width=\textwidth]{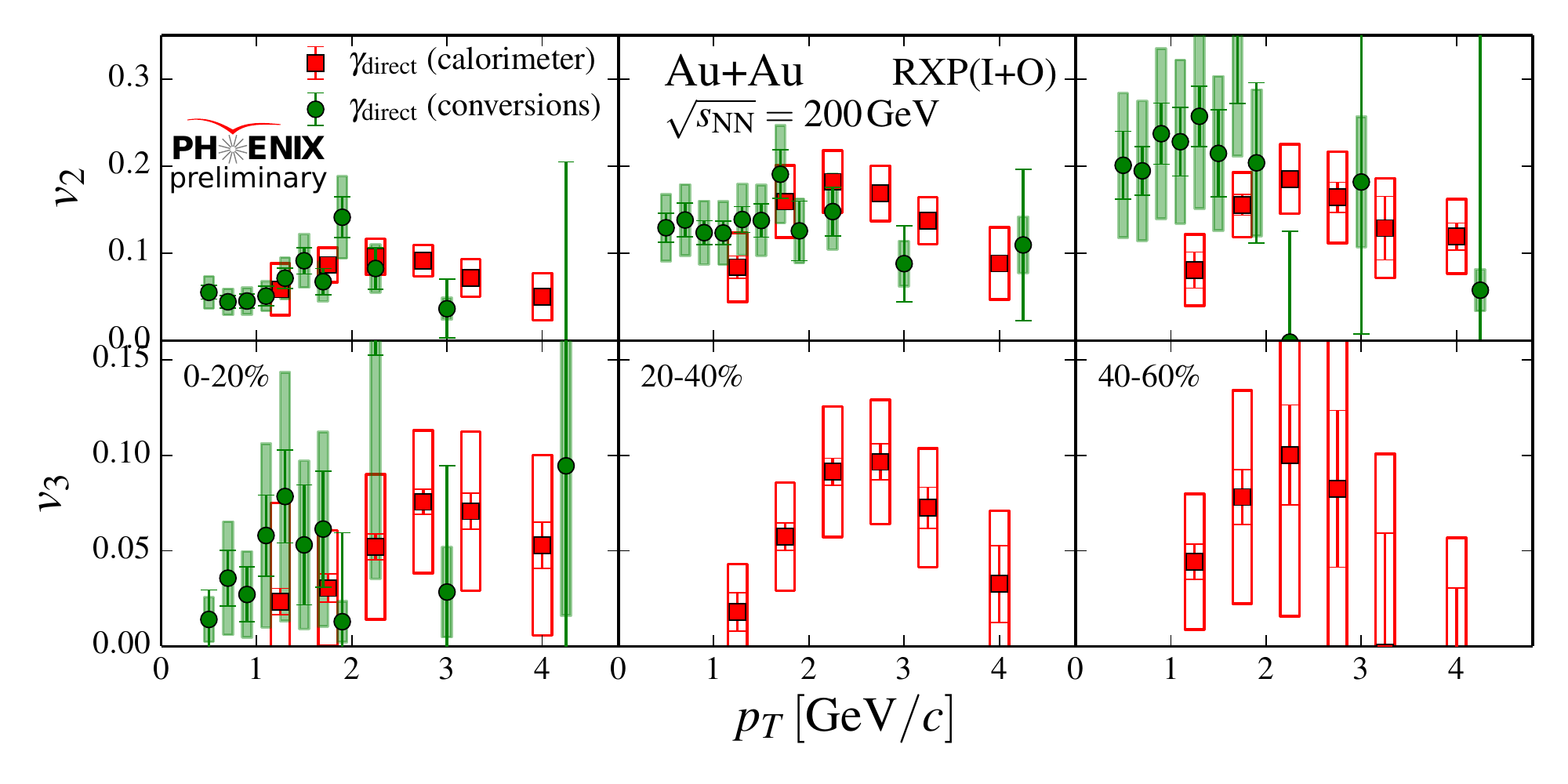}}
\end{minipage}
\caption{a) Integrated thermal photon yields as a function of $N_{part}$ for different lower $p_T$ integration limits. The dashed lines are independent fits to a power-law. b) Azimuthal anisotropy $v_2$ and $v_3$ of direct photons, for three different centrality selections.}\label{fig:photon_int_yield}
\end{figure}

Earlier PHENIX measurements established a surprising behavior of direct photon production and elliptic flow at low transverse momenta~\cite{dirgam_old}. Current theoretical models cannot explain this so-called direct photon puzzle, a large excess of direct photon production over the $p+p$ baseline in Au+Au collisions at $\sqrt{s_{\rm NN}} = 200$ GeV, together with a large azimuthal anisotropy of direct photons.\\
A new measurement, using data taken in 2010, offers a significantly improved precision for both the direct photon spectra and the azimuthal anisotrop~\cite{thermal_gamma}. This new measurement uses the conversion of real photons in the detector material for the measurement, it confirms the previously published data and extends the transverse momentum range towards lower $p_{T}$. 
The new data furthermore allow the analysis in finer centrality bins than before. The thermal photon yield is extracted as the excess of the direct photon production over the binary-scaled $p+p$ baseline. It is fit with an exponential, the slopes of these functions are independent of the centrality selection within the uncertainties of the measurement, with an average slope of $\sim$240 MeV/$c$.\\

The integrated thermal photon yield has been calculated in four centrality bins for different $p_{T}$ ranges, it is shown in Figure~\ref{fig:photon_int_yield}a as a function of $N_{part}$. We observe a scaling of the integrated yield with a power law function $AN^{\alpha}_{part}$. The exponent $\alpha$ was found to be common for all $p_{T}$ integration ranges, with $\alpha = 1.48 \pm 0.08\rm{(stat.)} \pm 0.04\rm{(syst.)}$. This is in the range of recent theoretical models describing direct photon emission in Au+Au collisions at this energy~(see references in \cite{thermal_gamma}).\\
The new methods have also been used to measure the direct photon azimuthal anisotropy, for the first time, the triangular flow $v_3$ has also been measured. The $p_T$ range and the precision of $v_2$ have been improved compared to the previously published data. The direct photon $v_2$ and $v_3$ are shown in Figure~\ref{fig:photon_int_yield}b. The result on both $v_2$ and $v_3$ puts strong new constraints on the modeling of the hydrodynamic time evolution and the modeling of radiation emission in heavy ion collisions.

\section{Quarkonia and Heavy Flavor}

\begin{figure}[ht]
\begin{minipage}[h]{65mm}
\centerline{\includegraphics[width=\textwidth]{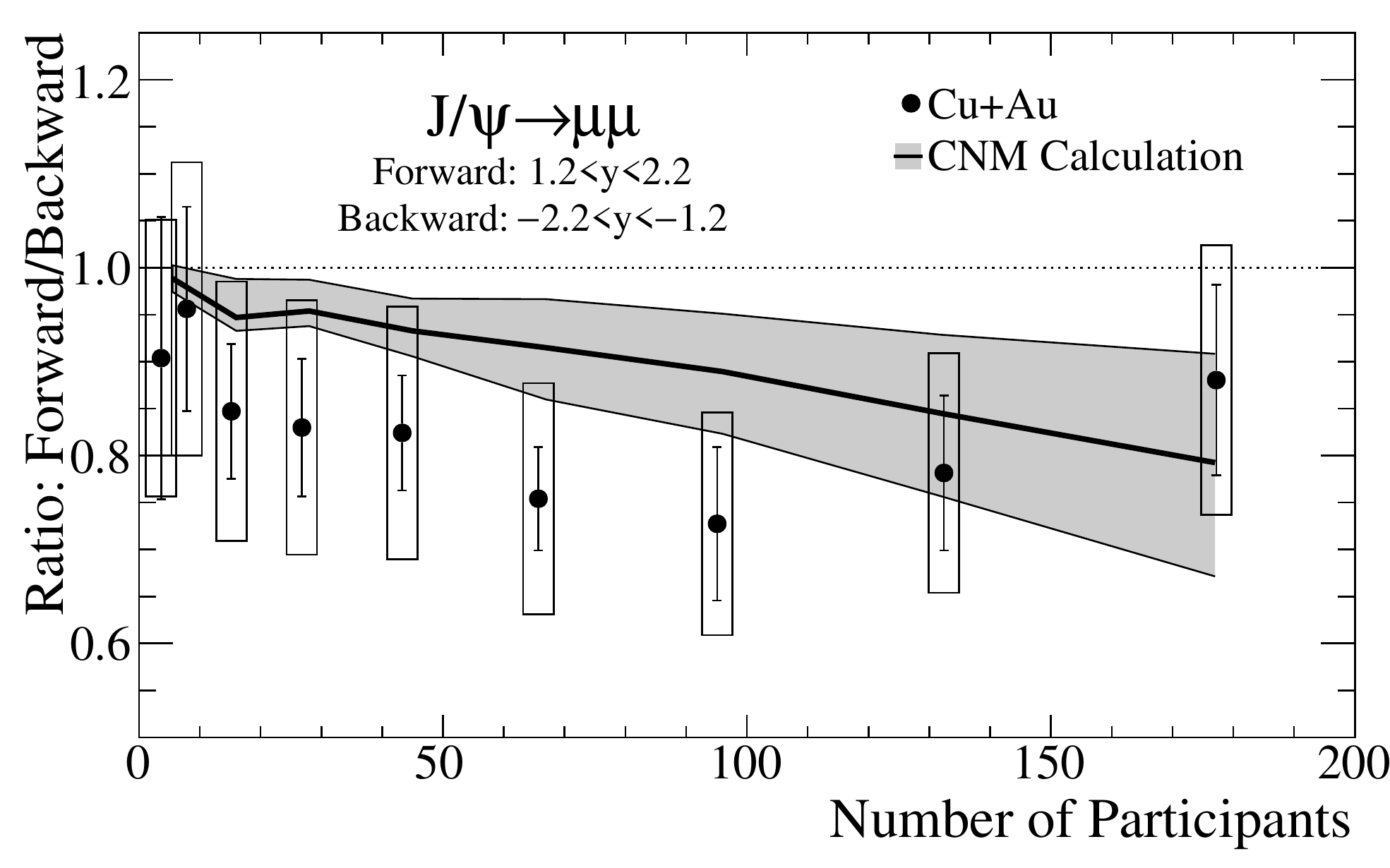}}
\end{minipage}
\hspace{\fill}
\begin{minipage}[h]{65mm}
\centerline{\includegraphics[width=\textwidth]{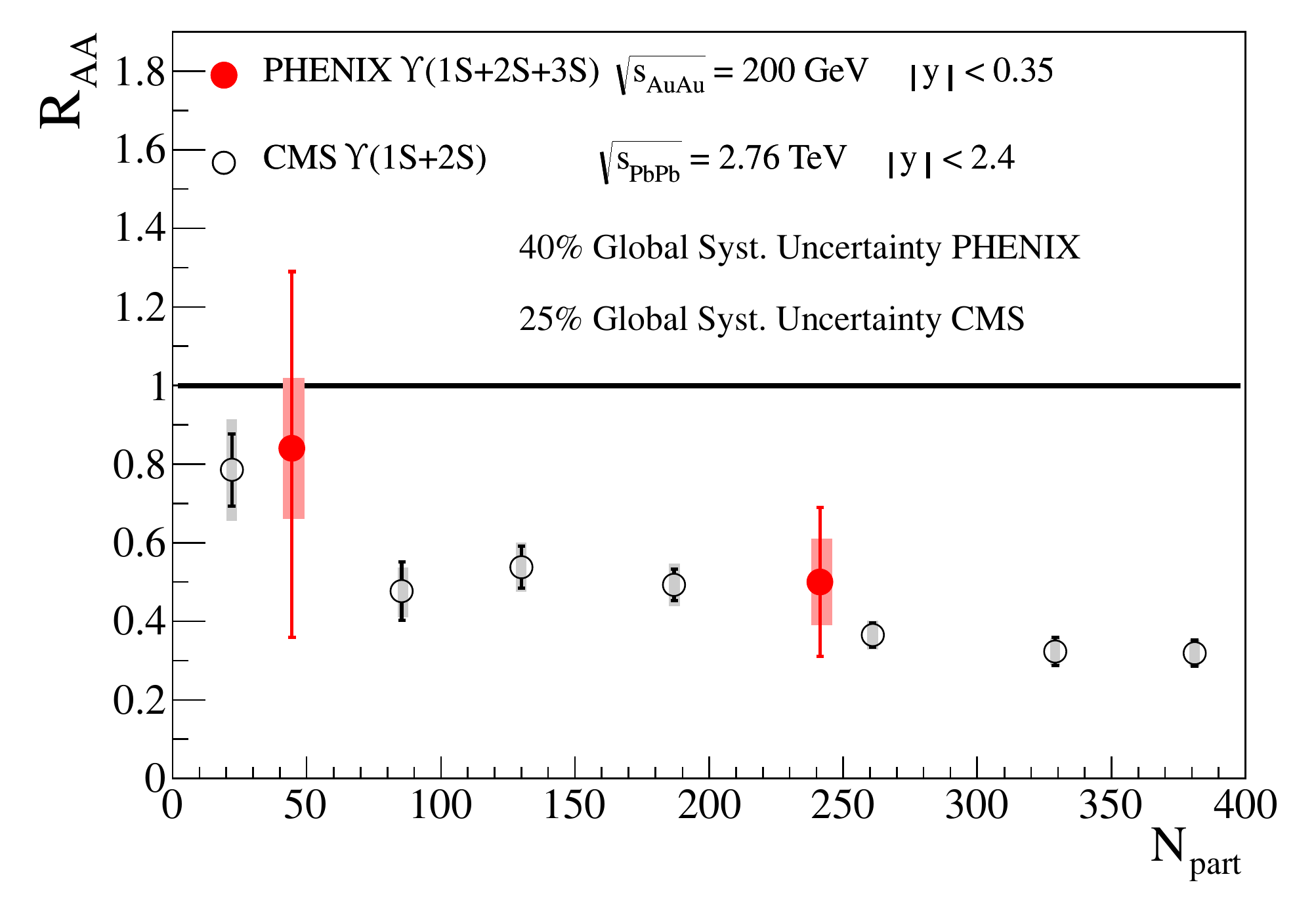}}
\end{minipage}
\caption{a) Ratio of forward- and backward-rapidity (Cu-going/Au-going) $J/\psi$ yields measured in Cu+Au collisions, with a model calculation for the contributions from cold nuclear matter~(see \cite{jpsi} and references therein). b) $\Upsilon$ nuclear modufication factor $R_{\rm AA}$ in Au+Au collisions at $\sqrt{s_{\rm NN}}$ = 200 GeV~\cite{upsilon}, compared to CMS results at LHC energies.}\label{fig:quarkonia}
\end{figure}

Heavy quarks are produced predominantly in hard scattering processes in the early phase of the collisions and hence can be used as a probe for the evolution of the medium. Quarkonia can be used like a thermometer for the medium, for example the $\Upsilon$ dissolves at a lower temperature than the $J/\psi$. PHENIX has excellent capabilities to measure heavy flavor production and quarkonia through leptonic channels.\\
One result of the $J/\psi$ measurement in Cu+Au collisions at $\sqrt{s_{\rm NN}}$ = 200 GeV is the ratio of the yields in the Cu-going direction and the Au-going direction~\cite{jpsi}. This ratio is shown in Figure~\ref{fig:quarkonia}a as a function of centrality, represented by the number of participants. The suppression of the $J/\psi$ is stronger in the Cu-going direction which is consistent with more low-x shadowing in the Au nucleus than the Cu nucleus. The result is also consistent with observations in $d$+Au collisions at the same energy where the suppression is stronger in the $d$ going direction.\\
The $\Upsilon$ has been measured in Au+Au collisions at the same energy~\cite{upsilon}. The resulting $R_{\rm AA}$ is shown in Figure~\ref{fig:quarkonia}b, a suppression of the $\Upsilon$ is seen in more central collisions. The suppression is consistent with the disappearance of the $2s$ and $3s$ states of the meson. A similar suppression within uncertainties can be seen in Pb+Pb collisions at the LHC.

\begin{figure}[ht]
\begin{minipage}[h]{70mm}
\centerline{\includegraphics[width=\textwidth]{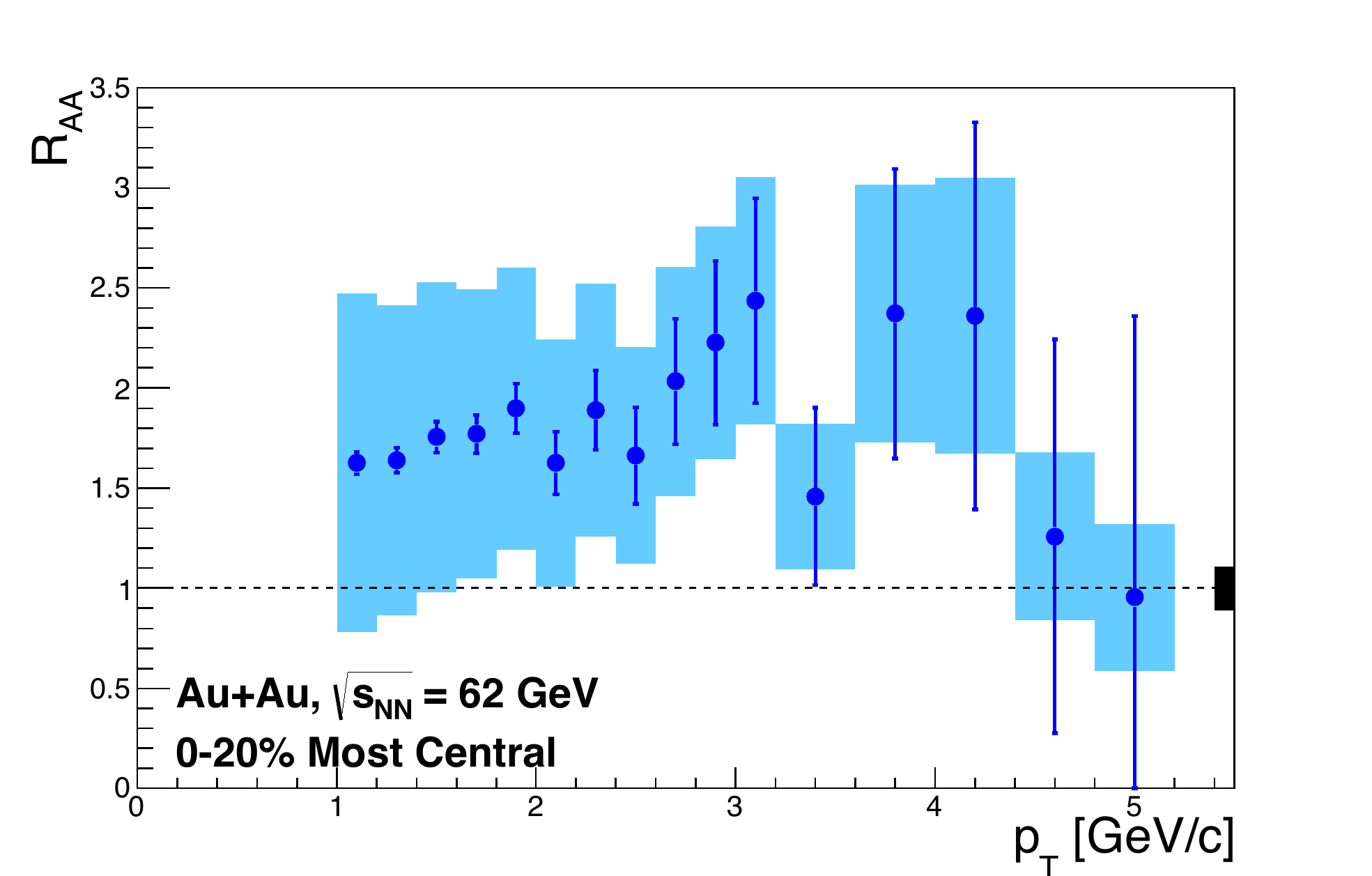}}
\end{minipage}
\hspace{\fill}
\begin{minipage}[h]{70mm}
\centerline{\includegraphics[width=\textwidth]{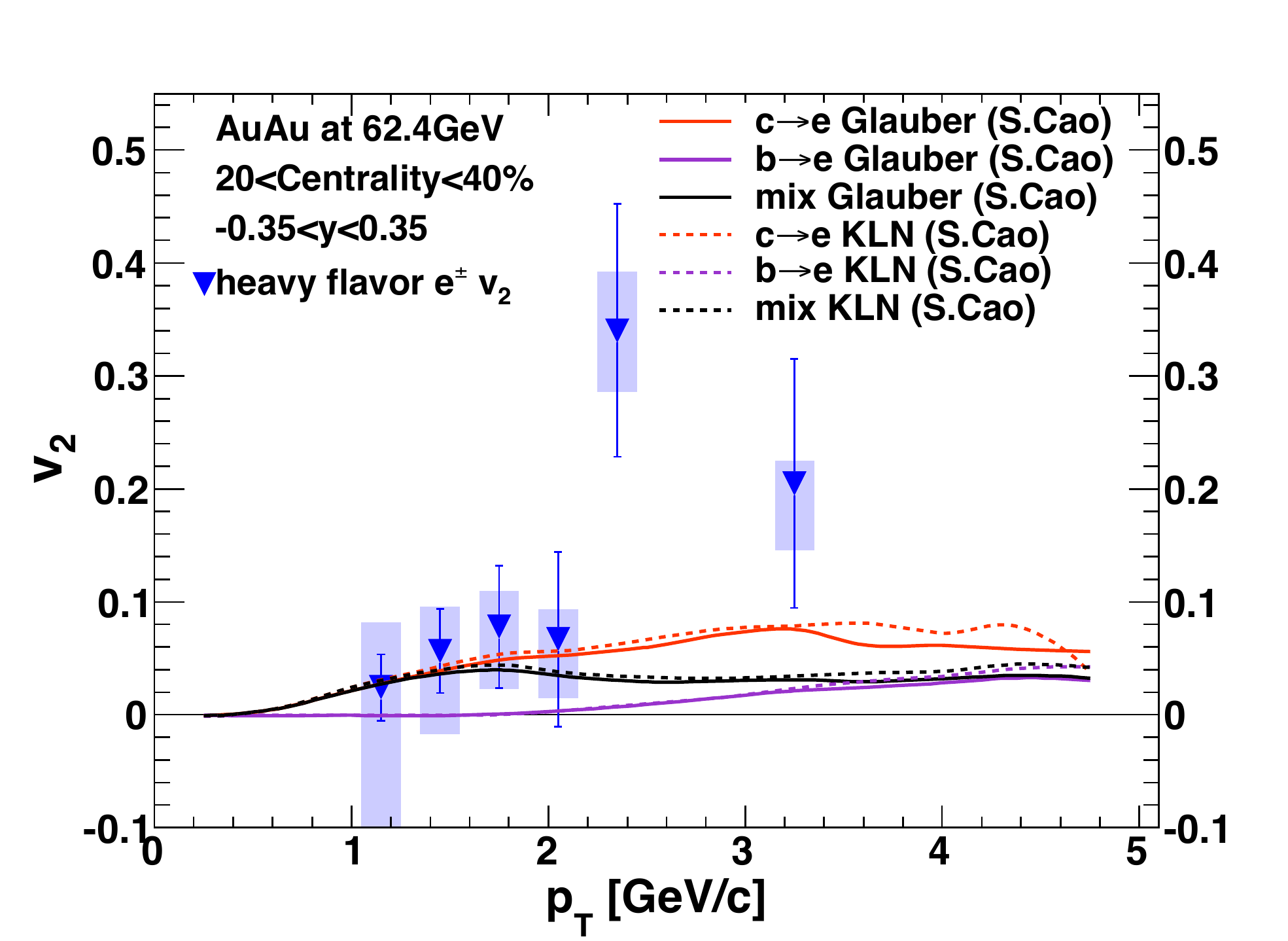}}
\end{minipage}
\caption{a) Nuclear modification factor $R_{\rm AA}$ of heavy-flavor electrons in 0-20\% central Au+Au collisions at $\sqrt{s_{\rm NN}}$ = 62.4 GeV. b) Heavy-flavor electron $v_2$, compared to model calculations, at the same energy~(see \cite{hf_62} and references therein)}\label{fig:heavy_flavor}.
\end{figure}
Heavy flavor production and its azimuthal anisotropy was measured via single electrons in Au+Au collisions at a lower energy of $\sqrt{s_{\rm NN}}$ = 62.4 GeV.~\cite{hf_62} Surprisingly, $R_{\rm AA}$ of these single electrons shows an enhancement over the $p+p$ reference at this energy, as seen in Figure~\ref{fig:heavy_flavor}a. This is opposite to the measurement of the same observable at  $\sqrt{s_{\rm NN}}$ = 200 GeV. The $v_2$ measurement, shown in Figure~\ref{fig:heavy_flavor}b, hints at a small but positive $v_2$ for heavy-flavor electrons which is also predicted by theoretical models. With the currently large uncertainties, no final conclusion can be drawn, hence it is important to improve the measurement with a new dataset that should be obtained in the future.
 
 \section{Initial State Effects}

\begin{figure}[ht]
\begin{minipage}[h]{64mm}
\centerline{\includegraphics[width=\textwidth]{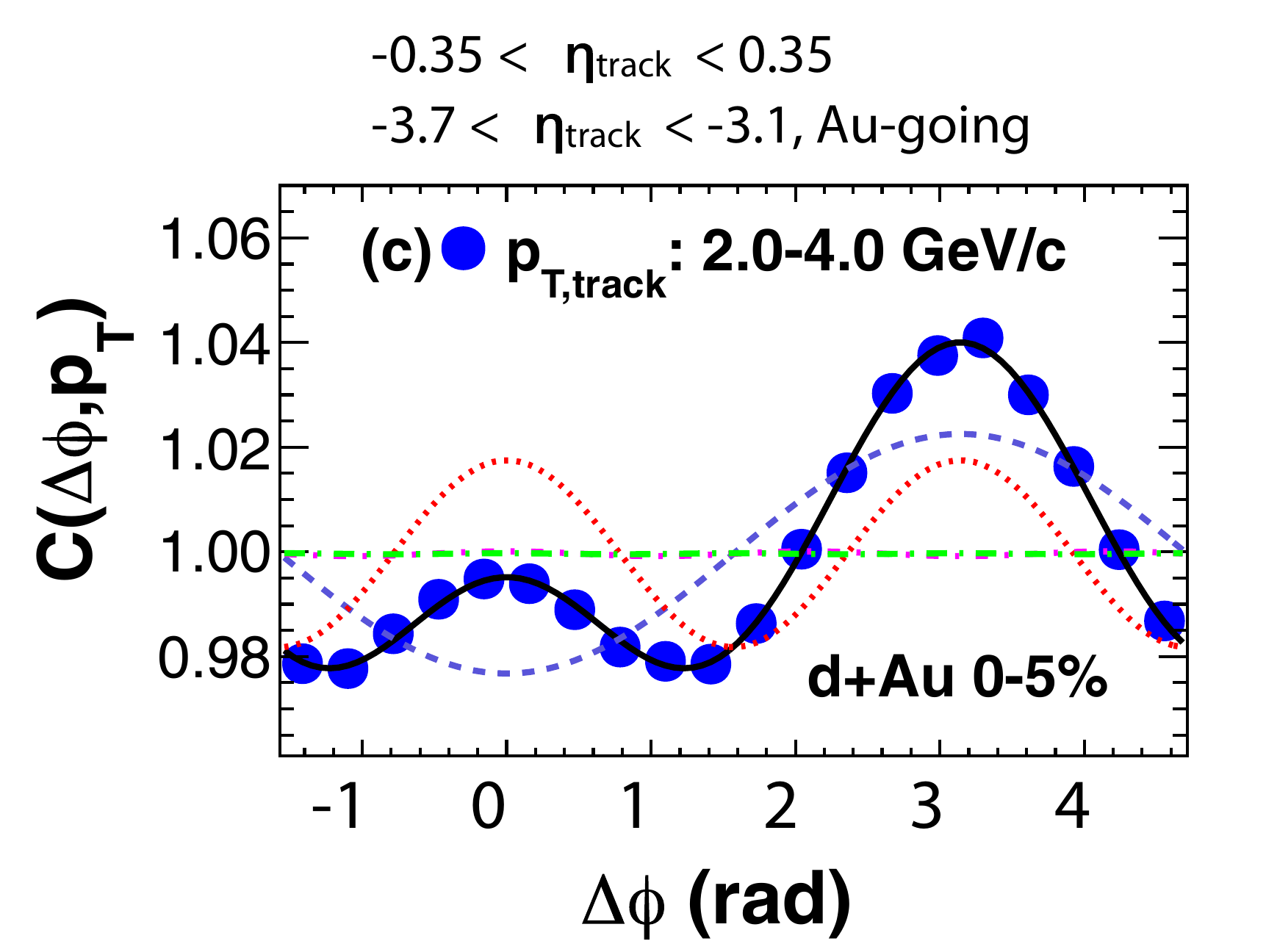}}
\end{minipage}
\hspace{\fill}
\begin{minipage}[h]{77mm}
\centerline{\includegraphics[width=\textwidth]{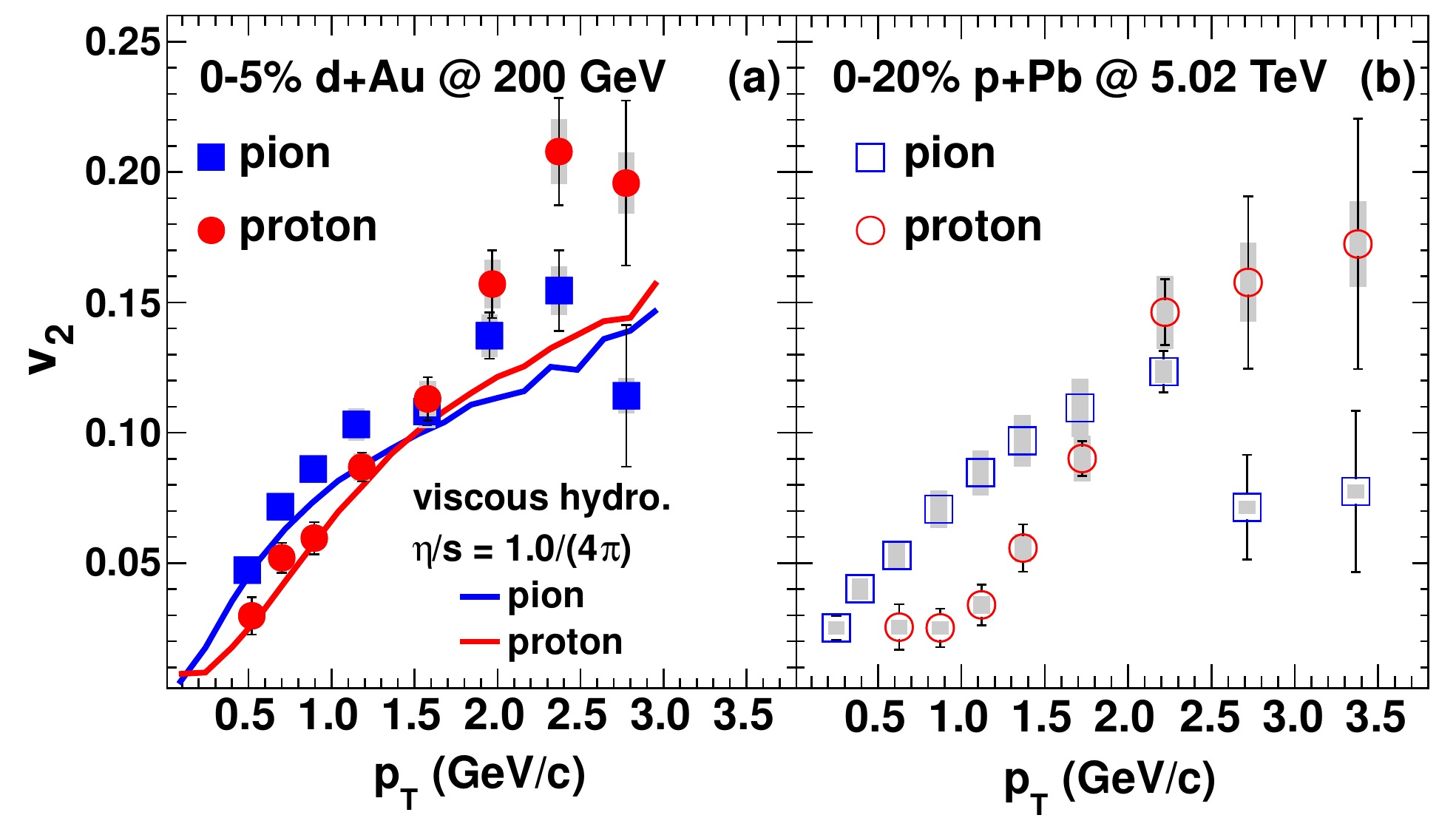}}
\end{minipage}
\caption{a) Azimuthal correlation function $C(\Delta \phi, p_T)$ for track-tower pairs in 0-5\% central $d$+Au collisions at $\sqrt{s_{\rm NN}}$ = 200 GeV, together with a four-term Fourier expansion fit. The individual components are drawn as well. b) $v_2$ of identified pions and (anti-)protons in the same collisions. The PHENIX data are compared with a hydrodynamic calculation in the left panel~\cite{model_dau}, and with LHC data for central $p$ + Pb collisions in the right panel of the figure~\cite{alicev2}.}\label{fig:azimuthal_corr}
\end{figure}

LHC analyses on $p$+Pb collisions at $\sqrt{s_{\rm NN}}$ = 5.02 TeV have indicated strong azimuthal long-range correlations of hadron pairs, PHENIX has complemented these results with a measurement of the charged particle $v_2$ in $d$+Au collisions at $\sqrt{s_{\rm NN}}$ = 200 GeV, using a small rapidity gap of $\Delta \eta$ = 0.47 to 0.7~\cite{dau_old}. So far, the reason for these anisotropies is now known.\\
A new analysis of the azimuthal angular correlations between charged hadrons at central rapidity and the energy deposited in a calorimeter at forward (Au-going direction) rapidity, with a pseudorapidity gap of $\Delta \eta > 2.75$, shows an enhanced near-side angular correlation~\cite{azimuthal_dau}. Figure~\ref{fig:azimuthal_corr}a~depicts this ridge-like correlation in central $d$+Au collisions. This result confirms the earlier PHENIX measurement with a smaller rapidity gap.\\
The azimuthal anisotropy $v_2$ has been measured as well with a large rapidity gap of $\Delta \eta > 2.75$ between the event plane and the observed particles. The measurement has been done with identified charged particles (pions and protons). The result of this measurement can be seen in Figure~\ref{fig:azimuthal_corr}b, compared with results in $p$+Pb collisions at the LHC. At both energies, the same mass ordering as in heavy-ion collisions is observed. The PHENIX data are compared to a calculation with Glauber initial conditions for viscous hydrodynamics with $\eta/s = 1.0/(4\pi)$, starting at $\tau = 0.5$ fm/$c$, followed by a hadronic cascade~\cite{model_dau}. The mass splitting at lower $p_T$ is seen in the calculation as well, the mass splitting is larger at the LHC which might indicate a larger radial flow in the higher energy regime.

\section{Future}

Currently, the PHENIX collaboration is working on the construction of a new detector for future RHIC runs. It will cover $2\pi$ in azimuth and include excellent capabilities to measure jets with electromagnetic and hadronic calorimetry. It is described in more detail in~\cite{sphenix}.

\begin{footnotesize}



%

\end{footnotesize}


\end{document}